\documentstyle[graphicx,psfig]{mn}

\def\spose#1{\hbox to 0pt{#1\hss}}

\def\multleft#1{\hbox to size{\vbox {\halign {\lft{##}\cr #1}}\hfill}\par}
\def\multright#1{\hbox to size{\vbox {\halign {\rt{##}\cr #1}}\hfill}\par}

\def\boxit#1{\vbox{\hrule\hbox{\vrule\kern3pt\vbox{\kern3pt
          #1 \kern3pt}\kern3pt\vrule}\hrule}}

\def\cm{{\rm\thinspace cm}}

\def\erg{{\rm\thinspace erg}}
\def\eV{{\rm\thinspace eV}}

\def\keV{{\rm\thinspace keV}}
\def\km{{\rm\thinspace km}}

\def\s{{\rm\thinspace s}}

\def\chisq{\hbox{$\chi^2$}}

\def\pcmcu{\hbox{$\cm^{-3}\,$}}

\def\ergpcmsqps{\hbox{$\erg\cm^{-2}\s^{-1}\,$}}

\def\ergps{\hbox{$\erg\s^{-1}\,$}}

\def\kmps{\hbox{$\km\s^{-1}\,$}}

\def\pcmsq{\hbox{$\cm^{-2}\,$}}

\title[Iron Line Spectroscopy of NGC~4593 with {\it XMM-Newton}]{Iron Line Spectroscopy of NGC~4593 with {\it XMM-Newton}:\\ 
Where is the Black Hole Accretion Disk?}

\author[C.~S.~Reynolds et al.]{
\parbox{15cm}{
Christopher~S.~Reynolds$^1$,
Laura~W.~Brenneman$^1$,
J\"orn~Wilms$^{2,3}$, and
Mary~Elizabeth~Kaiser$^4$}\\
$^1$Dept.\ of Astronomy, University of Maryland, College Park, MD 20742, USA.\\
$^2$Institut f\"ur Astronomie und Astrophysik, Abt.\ Astronomie, 
Universit\"at T\"ubingen, Sand 1, 72076 T\"ubingen, Germany\\
$^3$Department of Physics, University of Warwick, Coventry, CV4~7AL\\
$^4$Dept. of Physics and Astronomy, Johns Hopkins University, 3400 Charles Street, Baltimore, MD~21218.\\}

\date{In press}
\pagerange{\pageref{firstpage}--\pageref{lastpage}}
\pubyear{2003}

\begin{document}
\label{firstpage}
\maketitle

\begin{abstract}
  We present an analysis of the 2--10\,keV {\it XMM-Newton}/EPIC-pn
  spectrum of the Seyfert-1 galaxy NGC~4593.  Apart from the presence
  of two narrow emission lines corresponding to the K$\alpha$ lines of
  cold and hydrogen-like iron, this spectrum possesses a power-law
  form to within $\sim 3-5\%$.  There is a marked lack of spectral
  features from the relativistic regions of the black hole accretion
  disk.  We show that the data are, however, consistent with the
  presence of a radiatively-efficient accretion disk extending right
  down to the radius of marginal stability if it possesses low iron
  abundance, an appropriately ionized surface, a very high
  inclination, or a very centrally concentrated emission pattern (as
  has been observed during the Deep Minimum State of the Seyfert
  galaxy MCG--6-30-15).  Deeper observations of this source are
  required in order to validate or reject these models.
\end{abstract}


\section{Introduction}

The fluorescent K$\alpha$ emission line of iron is currently the best
probe we have to study strong-field gravitational effects in the
vicinity of black holes.  This line, together with an associated
backscattered continuum, is readily formed when a hard X-ray continuum
source irradiates the surface of a relatively cold and optically-thick
slab of gas (Basko 1978; Guilbert \& Rees 1988; Lightman \& White
1988; George \& Fabian 1991; Matt et al. 1991).  Nowadays, the hard
X-ray source is usually identified with thermal Comptonization from an
accretion disk corona, and the optically-thick structure as the
accretion disk itself (see Reynolds \& Nowak 2003 for a recent
review).  The diagnostic power of these spectral features lies in
investigations of their Doppler broadening and gravitational
redshifting (Fabian et al. 1989; Laor 1991).  The best example to date
of using these features to probe strong-field gravity is the
Seyfert-1.2 galaxy MCG--6-30-15 (Tanaka et al. 1995; Wilms et al.
2001; Fabian et al. 2002; Reynolds et al. 2004).  This object displays
a highly-broadened and skewed iron line that is strongly suggestive of
emission from an accretion disk reaching down to near the radius of
marginal stability for a rapidly-rotating black hole.  As yet, there
is no competing model that can explain, in detail, the iron line
feature in MCG--6-30-15.

However, it is an open question whether these relativistic spectral
features are generic in the spectra of various classes of active
galactic nuclei (AGN).  Nandra et al. (1997a) used data from the {\it
  ASCA} observatory to conclude that relativistically-broadened iron
lines are very common features in the X-ray spectra of Seyfert-1
nuclei.  On the other hand, {\it ASCA} found that these emission lines were
often weaker and/or narrower in the X-ray spectra of low-luminosity
AGN (e.g., Reynolds, Nowak \& Maloney 2000; Terashima et al. 2002),
high-luminosity AGN (Nandra et al. 1997b) and radio-loud AGN
(Sambruna, Eracleous \& Mushotzky 1999).  

Recent results from the European Photon Imaging Camera (EPIC) on broad
the {\it XMM-Newton} observatory have painted a more complex picture.
While {\it XMM-Newton} has, indeed, found undisputed cases of broad
iron lines in the Seyfert galaxies MCG--6-30-15 (Wilms et al. 2001,
Fabian et al. 2002), MCG--5-23-16 (Dewangan, Griffiths \& Schurch
2003), NGC~3516 (Turner et al. 2002), Mrk335 (Gondoin et al.  2002),
and Mrk766 (Pounds et al. 2003a), other Seyfert-1 galaxies
appear to show an absence of such features, with the best example to
date being NGC~5548 (Pounds et al. 2003b).  Clearly, the presence or
absence of a relativistic iron line depends upon currently unknown
factors and is not a simple function of AGN class.

Progress must be made by careful analysis of as many AGN X-ray spectra
as possible.  With this motivation, this {\it Paper} presents a
careful analysis of the hard-band {\it XMM-Newton} EPIC-pn spectrum of
the Seyfert-1 galaxy NGC~4593 ($z=0.009$).  Section~2 describes in
brief our observation and data reduction techniques.  Section~3
presents an analysis of the 2--10\,keV spectrum of NGC~4593 and
demonstrates a marked absence of a ``standard'' relativistic iron
line, a result which is discussed in more detail in Section~4.
Section~5 summarizes our principal conclusions.

\section{Observations and data reduction}

{\it XMM-Newton} observed NGC~4593 for a total of 76\,ksec during
orbit 465.  All instruments were operating during this observation.
The EPIC-pn was operated in its small-window mode to prevent photon
pile-up, using the medium-thick filter to avoid optical light
contamination.  The EPIC MOS-1 camera took data in the fast
uncompressed timing mode, and the MOS-2 camera operated in prime
partial W2 imaging mode.  Though the MOS results will not be discussed
further here, they mirrored the pn data within the expected errors of
calibration effects.  The average count rate for the pn instrument was
29.78\,ct\,s$^{-1}$.

The pipeline data were reprocessed using version 5.4.1 of the Science
Analysis Software, and the matching calibration files (CCF from 2003
January 14 via http://xmm.vilspa.esa.es/ccf).  From these, we rebuilt
the calibration index file using {\tt cifbuild}, and then reprocessed
the EPIC data using the {\tt emproc} and {\tt epproc} tasks.  Bad
pixels and cosmic ray spikes (narrow time filtering) were removed from
the events files via the {\b evselect} task within the SAS.  Circular
extraction regions 40\,arcsecs in radius were then defined both on and
off the source for generating source and background spectra,
respectively, using the {\tt xmmselect} task.  Response matrices and
ancillary response files were created using {\tt rmfgen} and {\tt
  arfgen}, and the data were then grouped with their respective
response files using the {\tt grppha} such as to impose a minimum of
25 counts per spectral bin.  This binning is required in order to get
sufficient counts per bin to make $\chisq$-spectral fitting a valid
statistical process.  Spectral modeling and analysis was performed
using the XSPEC package.

\section{The 2--10\,keV EPIC-pn spectrum}
\label{sec:results}

In this paper, we analyze only the hard band (2--10\,keV) data from
the EPIC-pn camera.  This energy restriction prevents us from having
to model the soft excess, ionized absorption, and recombination line
emission present in the 0.5--2\,keV band.  The soft spectrum of this
object will be discussed in detail in another publication (Brenneman
et al. 2004).

\begin{figure*}
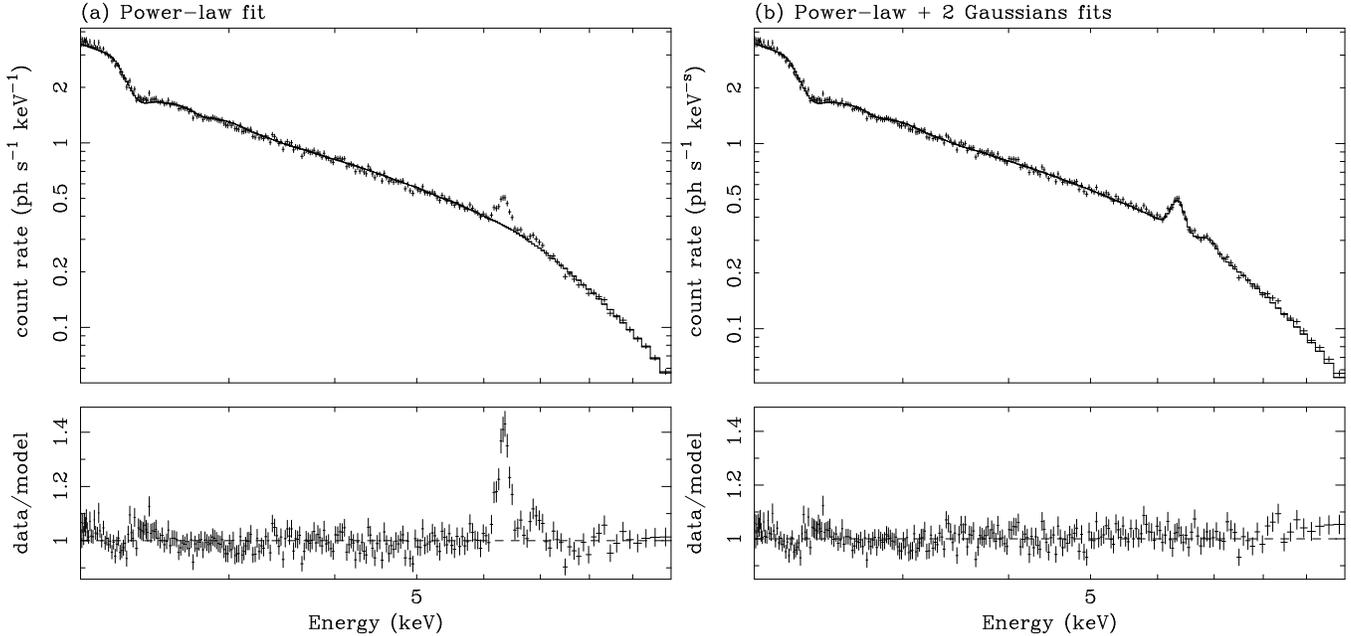

\hbox{
\psfig{figure=f1a.ps,width=0.5\textwidth,angle=270}
\psfig{figure=f1b.ps,width=0.5\textwidth,angle=270}
}
\caption{Spectral fits to the 2--10\,keV EPIC-pn data from {\it 
    XMM-Newton}.  Panel (a) shows the fit with a model consisting of a
  simple power-law modified by the effects of Galactic absorption
  ($N_{\rm H}=1.97\times 10^{20}\pcmsq$).  Two prominent emission
  lines are clearly visible in the lower ratio plot --- one
  corresponds to the K$\alpha$ line of cold iron (less than Fe\,{\sc
    xvii}) while the other corresponds to the K$\alpha$ line of
  hydrogen-like iron (Fe\,{\sc xxvi}).}
\label{fig:spectrum}
\end{figure*}

Initially, we fit this spectrum with a model consisting of a simple
power-law modified by neutral Galactic absorption with a column
density of $N_{\rm H}=1.97\times 10^{20}\pcmsq$ (Elvis, Wilkes \&
Lockman 1989).  The best fitting power-law index is $\Gamma=1.69\pm
0.01$ ($\chi^2/{\rm dof}=1993/1476$) giving a model 2--10\,keV flux
and luminosity of $4.41\times 10^{-11}\ergpcmsqps$ and $7.60\times
10^{42}\ergps$, respectively.  This best-fit is shown in
Fig.\ref{fig:spectrum}a.  The power-law describes the spectrum well
apart from two obvious line-like features between 6--7\,keV.  Modeling
these features with two Gaussian profiles improves the overall fit
significantly ($\chi^2/{\rm dof}=1526/1470$).  Best-fitting
(rest-frame) centroid energies, standard deviations and equivalent
widths for these two features are; $E_1=6.39\pm 0.01\keV$,
$\sigma_1=86\pm 17\eV$, EW$_1=117\pm 11\eV$ and $E_2=6.95\pm
0.05\keV$, $\sigma_2=102^{+94}_{-79}\eV$, EW$_2=39\pm 13\eV$.  It
seems very likely that these features are the K$\alpha$ emission lines
of ``cold'' iron (i.e., less than Fe\,{\sc xvii}) and hydrogen-like
iron that are expected at 6.40\,keV and 6.97\,keV, respectively.  We
do not detect a helium-like iron line at 6.67\,keV --- the upper limit
to the equivalent width of any such line is EW$<13\eV$.

\subsection{The narrow emission lines}

These relatively narrow emission lines likely originate from material
that is rather distant from the central black hole.  The cold iron
line is centered on the systemic velocity of NGC~4593 and is well
resolved (FWHM=$10900\pm 2200\kmps$).  For comparison, the broad
optical H$\beta$ line in NGC~4593 has FWHM=$4910\pm 300\kmps$ (Grupe
et al. 2004), approximately half of the velocity width of the cold
iron line.  Thus, it seems clear that the cold iron line is
originating from a region that lies significantly inside of the
optical broad emission line region (OBLR), and hence cannot be
identified with X-ray reflection from the classic ``molecular gas
torus'' postulated by unified Seyfert schemes.  We note that the {\it
  XMM-Newton} data sets no useful constraints on the presence of a
Compton backscattered continuum expected from X-ray reflection by cold
matter, thus the possibility remains that this line might be formed by
the X-ray illumination and subsequent fluorescence of optically-thin
material (as opposed to the optically-thick material normally
envisaged in X-ray reflection).

The hydrogen-like iron line, on the other hand, is only marginally
resolved (FWHM=$12200^{+11200}_{-9400}\kmps$).  Thus, it is not
possible to conclude where, relative to the OBLR, the ionized emission
originates.  It is, however, possible to say something about the
physical process underlying this emission.  If one supposes that this
line is emitted by collisionally-ionized thermal plasma (described by
the {\tt mekal} model in XSPEC; Mewe, Gronenschild \& van~den~Oord
1985; Mewe, Lemen \& van~den~Oord 1986; Kaastra 1992; Liedahl,
Osterheld \& Goldstein 1995), the EPIC data demand that the plasma
possess a temperature of at least $kT\sim 50\,keV$ in order to
reproduce the constraint on the hydrogen-like/helium-like equivalent
width while simultaneously not curving the overall continuum
excessively.  The required emission measure would be $EM\equiv
n^2V=1.6\times 10^{67}\pcmcu$, where $n$ is the electron number
density and $V$ is the volume of the thermal plasma.  If we further
suppose that this plasma surrounds the central engine in a spherical
geometry which is optically-thin to Compton scattering (or else we
would not observe rapid X-ray variability from the AGN), we can use
the column density and emission measure constraints to conclude that
the thermal plasma must have an extent of at least $4\times
10^{16}\cm$.  It is hard to understand the existence of such high
temperature plasma many thousands of gravitational radii from the
black hole.  Thus we disfavour this origin for the ionized iron line.

It is more likely that the ionized iron emission line originates via
radiative recombination and resonant scattering in strongly
photoionized gas within the central engine of the AGN.  While a
detailed exploration of this is beyond the scope of this paper, it is
tempting to identify this feature with the same plasma that produces
the highly ionized absorption features seen in several other AGN
(Reeves, O'Brien \& Ward 2003; Pounds et al. 2003).

\subsection{Limits on a broad iron line}

Is there evidence for a relativistically-broad iron line once the
narrow lines have been modeled?  To answer this question, we add a
relativistic iron line to the spectral fit using two models to
describe its profile; the Schwarzschild model of Fabian et al. (1989)
as implemented in the {\tt diskline} model of {\sc xspec} package, and
the near-extreme Kerr model of Laor (1991) as implemented in {\sc
  xspec}'s {\tt laor} model.  The energy of the emission line, $E_{\rm
  broad}$, was allowed to vary across the range of possible
Fe~K$\alpha$ transition energies, 6.40\,keV to 6.97\,keV (rest-frame).
The inner radius of the emitting region $r_{\rm in}$, the emissivity
index\footnote{The emissivity index is defined such that the
  emissivity of the line per unit proper area of the accretion disk is
  proportional to $r^{-\beta}$, where $r$ is the Boyer-Lindquist
  radius.} of the disk $\beta$, the viewing inclination $i$, and the
line normalization are also free parameters in the fit.  The outer
radius of the line emitting region was fixed at $r=1000r_g$ (where
$r_g=GM/c^2$).  The improvement in the goodness of fit was
$\Delta\chi^2=11$ and $\Delta\chi^2=10$ (for 5 additional degrees of
freedom) for the {\tt diskline} and {\tt laor} models, respectively.
Such a change is {\it not} significant at the 90\% level.

Thus, we have not obtained a detection of a broad iron emission line
in the EPIC-pn spectrum of NGC~4593.  In order to obtain a meaningful
upper limit to the equivalent width of any broad iron line, we must
specify its shape (since the data is incapable of doing that itself).
We can proceed either empirically or theoretically.  Empirically, we
can assume that any such line has the ``typical'' profile found in
co-added {\it ASCA} data by Nandra et al. (1997a), i.e., the {\tt
diskline} model with $E_{\rm broad}=6.4\keV$, $r_{\rm in}=6r_{\rm g}$,
$\beta=2.5$, $i=29^\circ$.  Using these assumptions, we can set an
upper limit (with a 90\% confidence level for one interesting
parameter) to the broad line equivalent width of $W_{\rm
broad}=87\eV$.  Theoretically, the simplest model (Shakura \& Sunyaev
1973; Novikov \& Thorne 1974; Page \& Thorne 1974) is one in which the
accretion disk is geometrically-thin and radiatively-efficient,
extends from the radius of marginal stability to large radii, and with
an iron line emissivity that tracks the underlying dissipation
(Reynolds \& Nowak 2003; Reynolds et al.  2003).  Applying such a line
profile to the NGC~4593 data in the case of a near-extreme Kerr black
hole (with dimensionless spin parameter $a=0.998$) results in an upper
limit to the equivalent width of $99\eV$.  These are significantly
less than the values expected from theoretical reflection models
($\sim 200\eV$ for solar abundances; e.g., Matt, Fabian \& Reynolds
1997 and references therein) or observed in the Seyfert galaxy
MCG--6-30-15 ($\sim 400\eV$; Fabian et al. 2002).  Thus, there appears
to be a significant absence of spectral features from a relativistic
accretion disk.

\section{Where are the accretion disk signatures?}

As discussed in the Introduction, the black hole accretion paradigm of
AGN is very well established and supported by a significant body of
evidence.  Thus, the results of Section~\ref{sec:results} beg us to
turn the question around: why are we not seeing the X-ray reflection
signatures of a relativistic accretion disk, given that we believe
such a disk exists and is responsible for all of the AGN emissions
that we observe?  A straightforward solution to this problem would be
to hypothesize sub-solar abundances of iron in the black hole
accretion disk.  If the light elements are present in cosmic abundance
(Anders \& Grevesse 1989), one needs iron to be under-abundant by a
factor of 3 (i.e., $Z_{\rm Fe}< 0.3$) in order to reduce the
(cold) broad fluorescent emission below the 100\,eV level as required
by our data (e.g., see Reynolds, Fabian \& Inoue 1995).  If the light
elements are {\it overabundant}, the enhanced photoelectric absorption
further decreases the iron line equivalent width.  A light element
overabundance by a factor of two reduces the required iron
underabundance to only 30\% (i.e., $Z_{\rm Fe}< 0.7$).  

However, it would be surprising if the solution to the lack of a broad
iron line was simply an underabundance of iron, given the highly
evolved nature of stellar populations in the nuclei of galaxies such
as NGC~4593.  With this motivation, we explore modifications of the
``canonical'' line models discussed above and show that it is, in
fact, rather easy to bury relativistic spectral features in the noise
even if they are present at the level associated with cosmic abundance
material.

\subsection{Very broad lines and torqued accretion disks}

Wilms et al. (2001) and Reynolds et al. (2004) have analyzed the
EPIC-pn spectrum of MCG--6-30-15 in its Deep Minimum state and found
it to possess {\it very} broadened X-ray reflection features.  On the
basis of these data, they suggest that the accretion disk in
MCG--6-30-15 is being torqued by interactions with the central
spinning black hole, producing a dissipation that is very centrally
concentrated.  By employing the disk models of Agol \& Krolik (2001),
Reynolds et al. (2004) suggest that the Deep Minimum state of
MCG--6-30-15 corresponds to a torque-dominated
(``infinite-efficiency''; see Agol \& Krolik 2001) accretion disk
viewed at an inclination of $30-40^\circ$.

There is little evidence for an obscuring molecular torus in
MCG--6-30-15 (e.g., see Lee et al. 2002) and so, in principle, the
central accretion disk could be viewed at any angle.  It is
interesting to note that if MCG--6-30-15 were observed at a high
inclination with only moderate signal-to-noise, the Deep Minimum iron
line would be so broad as to be undetectable against the noise of the
continuum.

To explore whether this is also the case in NGC~4593, we add to the
spectral fit a cold iron line with a profile corresponding to an
infinite-efficiency accretion disk around a near-extreme Kerr black
hole ($a=0.998$).  The inclination and normalization of the line were
left as free parameters.  This leads to only a slight improvement in
the goodness of fit over the simple power-law plus narrow line model
($\Delta\chi^2=3$ for 2 additional degrees of freedom).  Since this
line is so broad, the upper limit on the equivalent width is $W_{\rm
  broad}<250\eV$.  Thus, a broad line with the strength expected from
a cosmic abundance accretion disk is consistent with these data if the
emissivity profile is very centrally concentrated.

\subsection{Highly-ionized accretion disks}

\begin{figure}
\centerline{
\psfig{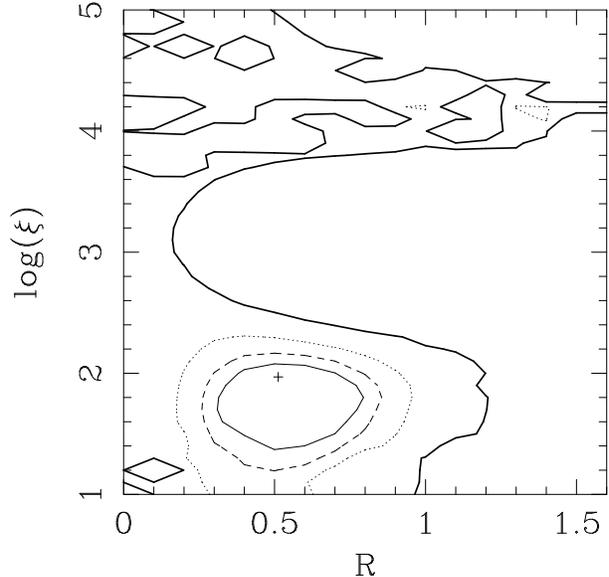}
}
\caption{Confidence contours on the $(\xi,{\cal R})$-plane.   Shown 
  here are the 68\% (thin solid line), 90\% (dashed line) and 95\%
  (dotted line) confidence contours using the best-fit ionized disk
  model (denoted by a cross).  Also shown (thick solid line) is the
  contour corresponding to a goodness of fit parameter ($\chi^2$)
  equal to the best fit power-law plus 2 narrow Gaussians.  One can
  see that a ``complete'' disk (i.e. ${\cal R}=1$) is allowed if
  either moderately ionized disks ($\log\xi\sim 2$) or highly ionized
  disks ($\log\xi\sim 4$).}
\label{fig:ionization}
\end{figure}

The prominence of X-ray reflection features can be significantly
reduced by ionization of the accretion disk surface.  Can ionization
explain the lack of inner disk features in the spectrum of NGC~4593?

We address this question using the ionized reflection models of
Ballantyne, Ross \& Fabian (2001) convolved with the relativistic
smearing model corresponding to a ``standard'' Novikov-Page-Thorne
accretion disk around a near-extremal Kerr black hole.
Figure~\ref{fig:ionization} shows confidence contours on the
$(\xi,{\cal R})$-plane, where $\xi$ is the ionization parameter of the
disk surface and ${\cal R}$ is the relative normalization of the disk
reflection spectrum (defined such that an isotropic source above a
plane-parallel reflector gives ${\cal R}=1$).  Ionized disk reflection
[with $\log(\xi)=2, {\cal R}=0.5$] provides the best fit of any of the
models reported in this paper ($\chi^2=1500/1470$) since it describes
subtle concaved curvature that is actually present in the data.  Using
this as our reference fit, it can be seen that an almost complete disk
(i.e. ${\cal R}$ almost unity) is possible provided the disk surface
has an ionization parameter in the range $\log\xi=1.5-2.5$.  In this
ionization range, the iron fluorescence line is strongly suppressed by
resonant scattering followed by Auger destruction (Matt, Fabian \&
Ross 1993) thereby allowing the absence of a broad fluorescence line
to be compatible with a complete disk.  Using the slightly looser
constraint that the ionized disk fit should be no worse than the
no-disk fit (i.e., $\chi^2<1526$; thick contour on
Fig.~\ref{fig:ionization}), we see that there is an additional
``finger'' of allowed parameter space with $\log(\xi)\approx 4$,
corresponding to highly ionized disks in which the iron atoms are
mostly fully stripped.

Thus, in conclusion, even a standard Novikov-Page-Thorne accretion
disk is permissible in this object as long as it is either highly
ionized (in which case the iron atoms are fully stripped) or
moderately ionized (when the fluorescent line is strongly suppressed
by resonant scattering and Auger destruction).

\subsection{Limb-darkening and coronal attenuation from a high-inclination 
disk}

As discussed in the Introduction, it is believed that any broad iron
line is produced via fluorescence in the outer few Thomson depths of
the optically-thick part of the black hole accretion disk.  If viewed
at high inclination (i.e., almost edge-on), there are two distinct
``limb-darkening'' effects that suppress the observed line flux (e.g.,
Reynolds. Maloney \& Nowak 2000).  Firstly, iron line photons can be
photoelectrically absorbed on their passage through the outer layers
of the disk, either by the K-shell edges of elements lighter than iron
or the L-shell edges of iron.  The resulting excited ion will then
de-excite either via the Auger effect or through the emission of new
fluorescent line photons.  Either way, the result is a net loss of
iron line photons from the observed spectrum.  Secondly, iron line
photons can be Compton scattered by energetic electrons in the X-ray
emitting corona.  If the corona can be described as a thermal plasma
with temperature $T$, the average fractional energy shift per Compton
scattering event is $\Delta E/E=4kT/m_ec^2$, which is of the order of
unity for typical coronal temperatures $kT\sim 100\keV$.  Thus,
Compton scattering effectively removes photons from the iron line,
spreading them out in energy space into a broad pseudo-continuum.

Both of these effects, photoelectric absorption in the disk and
Compton scattering in the corona, are strongly accentuated for high
inclination observers since the photons then have to follow
trajectories that graze the disk atmosphere and corona.  The primary
continuum photons, on the other hand, are not subject to
limb-darkening since the corona is optically-thin.  Hence, the
equivalent width of a broad iron line would be expected to drop
(possibly dramatically) as the disk is viewed increasingly edge-on.
While there is little evidence (e.g., excess neutral absorption) that
the inner disk of NGC~4593 is viewed at high-inclination, the
possibility that limb-darkening is responsible for the absence of a
broad line in this Seyfert nucleus cannot be ruled out with current
data.  If this is the correct explanation, a deeper observation might
be able to uncover the limb-darkened broad iron line displaying a
profile that has been modified by the effects of limb-darkening
(Beckwith \& Done 2004).

\subsection{Other possibilities}

Until now, we have been examining models of ``complete'' accretion
disks, i.e., disks that remain geometrically-thin and
radiatively-efficient from the radius of marginal stability to very
large radii.  In particular, we have been examining how extreme
broadening or ionization can render a reasonable strength iron line
undetectable, even in a moderately long {\it XMM-Newton} observation.
If one or both of these effects are responsible for the lack of an
obvious broad line in NGC~4593, higher signal-to-noise spectroscopy
either with a deeper {\it XMM-Newton} observation or (eventually) {\it
  Constellation-X} and {\it XEUS} will reveal the subtle signatures of
the relativistically smeared reflection.  However, it is also possible
that the broad iron line is genuinely absent.  There are two possible
scenarios.

\subsubsection{Advection dominated accretion disks}

Firstly, the accretion disk may possess a transition point inside of
which accretion switches to a hot, geometrically-thick, optically-thin
mode.  There has been extensive theoretical work on such flows (e.g.,
Ichimaru 1977; Rees 1982; Narayan \& Yi 1994; Blandford \& Begelman
1999; Narayan et al. 2002).  Of relevance here are the class of flows
that are advection-dominated (i.e., the locally-dissipated energy is
not locally radiated) because of their low accretion rate.  In terms
of their iron-K band spectra, these disks would lack any very broad
component, possessing a rather narrower iron line
resulting from fluorescence of the thin-disk surface beyond the
transition radius.  The equivalent width of these features is
determined by the solid angle that the thin-disk subtends at the X-ray
source (which, in this scenario, would be thermal bremsstrahlung and
Comptonization in the hot/geometrically-thick inner accretion flow).
Although dependent on the precise geometry, these iron K$\alpha$
equivalent widths can be reduced from the plane-parallel case by
factors of 2--4 (i.e. to 50--100\,eV).

Indeed, it is possible that we {\it are} seeing just such a component
in the form of the narrow 6.40\,keV line modeled in
Section~\ref{sec:results}.  If we model this feature with a {\tt
  diskline} profile possessing an inner truncation radius, instead of
a narrow Gaussian profile, we determine that the inner truncation
radius to the line emitting region is $r_{\rm in}>200r_{\rm g}$.  In
deriving this number, we have fitted the diskline model assuming an
emissivity index of $\beta=3$ (appropriate for illumination of a flat
disk by a central raised corona or advection dominated region), and a
rest-frame line energy of 6.40\,keV.  The equivalent width of this
feature is $W=113\pm 13\eV$ with a goodness of fit of $\chi^2/{\rm
  dof}=1529/1470$; i.e., the truncated diskline profile produces very
comparable results to the narrow Gaussian profile.

\subsubsection{Strongly beamed primary emission}

Secondly, it is possible that the observed X-ray continuum does not
originate from, and does not irradiate, the accretion disk.  The most
likely alternative is that the X-ray continuum originates from a
relativistic jet flowing from the inner regions; the primary X-rays
are then beamed away from the accretion disk, and X-ray reflection
from the disk is very strongly suppressed.  Given the high amplitude
variability seen on hour-timescales within this object, the X-rays
would have to originate from the inner parts of any jet.

\section{Conclusions}

In this paper, we have analyzed the 2--10\,keV {\it XMM-Newton}
EPIC-pn spectrum of NGC~4593, with particular emphasis on the
implications of this spectrum for the nature of the accretion flow in
this Seyfert-1 nucleus. Only two spectral features are detected which
can be identified with narrow K$\alpha$ emission lines of cold and
hydrogen-like iron at 6.40\,keV and 6.97\,keV, respectively.  Once
these have been modelled, the spectrum has a power-law form (to within
a 3-4\% accuracy) across the 2--10\,keV band.

We fail to detect any X-ray reflection features from a relativistic
accretion disk.  However, we show that, even if a
radiatively-efficient geometrically-thin complete accretion disk
exists, its X-ray reflection signatures would be buried in the noise
if the accretion disk has either a very centrally concentrated
irradiation profile or an appropriately ionized surface.  Either of
these models can be tested by longer EPIC observations which would be
sensitive to the subtle features displayed by a highly blurred or
ionized reflection spectrum.  If longer observations still
fail to detect any accretion disk signatures, we are forced to
consider other scenarios.  Firstly, the inner disk may be very hot and
optically-thin, thereby being incapable of producing any X-ray
reflection features. Secondly, the observed X-ray continuum might be
highly anisotropic, being strong beamed away from the disk (and
towards the observer) thereby rendering any disk features
undetectable.

\section*{Acknowledgments}

We thank Andrew Young for stimulating discussion throughout the course
of this work.  We acknowledge support from the NASA/{\it XMM-Newton}
Guest Observer Program under grant NAG5-10083, and the National
Science Foundation under grant AST0205990.

\end{document}